\documentclass[12pt,twoside]{article}
\usepackage{fleqn,espcrc1}
\input epsf
\newcommand{\ie}{{\em i.e.}}

\title{
\begin{flushright}
{\normalsize IIT-HEP-01/1\\
\vspace{ -.05in}
hep-ex/0101024
}
\end{flushright}
Prospects for low-energy antiproton physics at Fermilab\thanks{To 
appear in Proceedings of the LEAP 2000 Conference,
Venice, Italy, Aug.\ 20--26, 2000.}
}

\author{ Daniel M. Kaplan\thanks{E-mail: kaplan@fnal.gov}\address{Illinois 
Institute of Technology, \\
             Chicago, Illinois 60616, USA}  }  

\begin{document}
\maketitle

\begin{abstract}
Fermilab has long had the world's most intense antiproton source, but the
opportunities for medium-energy antiproton physics have been limited, and those
for low-energy antiproton physics nonexistent. The conclusion of E835 brings
this era to an end. While the future of antiproton physics at Fermilab remains
highly uncertain, developments are occurring that may lead to a low-energy
program within the next several years, with the possibility of an improved
medium-energy program thereafter. These issues were considered at the recent
$\overline{p}2000$ Workshop at Illinois Institute of Technology. I summarize
the current status of the Fermilab antiproton facility, review hyperon {\em CP}
violation as an example of the physics that might be achievable, and discuss
future possibilities.
\end{abstract}

\section{{\boldmath $\overline{p}2000$} WORKSHOP: FUTURE OF 
{\boldmath $\overline{p}$}s AT FERMILAB}

The $\overline{p}2000$ Workshop was held Aug.~3--5, 2000 at Illinois Institute
of Technology. Its goal was to explore the physics
potential of a possible new low-energy antiproton storage ring at Fermilab and
begin the process of planning for its construction and exploitation.
Thirty-nine physicists attended from Europe, Asia, and the U.S., and promising
ideas were presented.  The major questions addressed
could be summarized as follows:

\begin{enumerate}
\item
What will be Fermilab's capabilities in $\overline{p}$ 
physics during the next several years?

\item
Can a strong physics program be identified to take advantage of these 
capabilites?

\end{enumerate}

\subsection{Physics motivations}

In particular, we considered two motivating ideas and 
four general physics areas in which significant progress
might be made in a program at Fermilab:

\begin{itemize}

\item
Fermilab is now starting to define its directions for the post-LHC era. Further
broadening of the non-``energy frontier" physics program (beyond BTeV and
the neutrino
and Main Injector fixed-target efforts) could be an appealing option if
a strong clientele can be found.

\item 
Fermilab already has the world's highest-intensity antiproton source, and the
available $\overline{p}$ intensity will continue to increase.

\item
The physics reach in $\overline{p}p\to\,$hyperons might benefit from greater 
luminosity.

\item The physics reach in $\overline{p}p\to\,$charmonium might benefit from 
greater luminosity, higher energy, and improved energy precision.

\item
The physics reach in $\overline{p}p\to\,$glueballs or gluonic hybrids  might
benefit from  more running and higher energy.

\item 
A facility for decelerating and trapping antiprotons at Fermilab could provide
much more beam than the AD.

\end{itemize}

\subsection{Current capabilities}

The $\overline{p}$ source now produces cooled antiprotons at a maximum 
``stacking rate"
of $\approx$10\,mA/hr.  Given the circumference of the antiproton accumulator, this
corresponds to a production rate of $10^{11}\,$antiprotons/hr and can
support a maximum luminosity of about $2\times10^{32}\,{\rm cm}^{-2}{\rm
s}^{-1}$, \ie\ beyond this luminosity collisions would consume antiprotons
faster than they are produced.  The goals for Tevatron Run II are 20\,mA/hr by
March, 2001, and 100\,mA/hr by Run IIB ($\approx2005$).

The antiproton fixed-target run of Experiment 835 (charmonium
production) ends in November, 2000, after which no further
experimental running in the accumulator is expected. Once Run II begins, the
accumulator is expected to be fully dedicated to supplying antiprotons for the
Tevatron Collider, in support of two (and, once BTeV is running, possibly
three) 2\,TeV collision regions operating simultaneously at
$2\times10^{32}\,{\rm cm}^{-2}{\rm s}^{-1}$.  Thus once Run II begins,
Fermilab's approved and existing capabilities in low-energy antiproton physics
will be nil.

\subsection{Possible low-energy-{\boldmath $\overline{p}$} upgrade}

In recent months Fermilab's G. Jackson has been exploring the possibility of
building a new, small, low-energy $\overline{p}$ facility at Fermilab. The
immediate motivations for such a facility come from fields outside 
nuclear and particle physics~\cite{Jackson,Jackson-private}:
NASA is pursuing ideas for
$\overline{p}p$-annihilation-fueled interstellar travel in the long-term future
and is interested in a source of trapped antiprotons to begin investigating
them.
Ideas have also been discussed for possible medical applications of trapped
antiprotons:
production of short-lived isotopes for positron-emission tomography and
use of antiprotons for cancer therapy.
Exploratory efforts to investigate provision of trapped
antiprotons for these purposes are going ahead, subject to the boundary
condition that they remain parasitic to commissioning and high-energy running
of the Tevatron Collider  and that they consume at most a few percent of the
available antiprotons. 

As discussed below, a stacking rate consistent with $10^{33}\,$cm$^{-2}$s$^{-1}$ luminosity may be
a desirable capability for a future facility; at least it is a plausible upper
limit. Such a high stacking rate is quite conceivable but may require upgrading
the proton source. It has been estimated~\cite{Chou} that the proposed 
``Proton Driver" upgrade~\cite{Proton-Driver} will increase the intensity in
the Main Injector by a factor $\approx$4. If other bottlenecks in the  
$\overline{p}$ production process can be overcome, this will translate into a
similar factor in stacking rate.

\section{A PHYSICS EXAMPLE}

To understand some of the issues for a future low-energy antiproton facility
(in particular, the need for $10^{33}$ luminosity), we consider in some detail 
a challenging
physics example: hyperon $C\!P$ violation.

\subsection{Hyperon {\boldmath $C\!P$} violation}

In addition to {\em CP} violation in kaon decays~\cite{Rosner}, the Standard
Model predicts a slight {\em CP} asymmetry in decays of
hyperons~\cite{Hyperon-CP,ACP,Valencia}.  The most accessible signals involve comparison
of the (nonuniform) angular distributions of the decay products of polarized
hyperons with those of the corresponding antihyperons~\cite{ACP}. For a
precision measurement, it is necessary to know the polarizations of the
initial hyperons and antihyperons to high accuracy.

By angular-momentum conservation, in the decay of a spin-1/2 hyperon to a
spin-1/2 baryon plus a pion, the final state must be either S-wave or P-wave.
As is well known, the interference term between the S- and P-wave decay
amplitudes gives rise to parity violation, parametrized by Lee and
Yang~\cite{Lee-Yang} in terms of two independent parameters $\alpha$ and
$\beta$: $\alpha$ is proportional to the real and $\beta$ to the imaginary part
of this interference term. {\em CP} violation can be sought as a difference in
$|\alpha|$ or $|\beta|$ for a hyperon decay and its {\em CP}-conjugate
antihyperon decay or as a particle-antiparticle difference in the partial
widths for such decays~\cite{ACP,Donoghue-etal}.

Table~\ref{tab:HCP} summarizes the experimental situation.  
The first three experiments cited studied
$\Lambda$ decay only~\cite{R608,DM2,PS185}, setting limits on the 
{\em CP}-asymmetry parameter~\cite{ACP,Donoghue-etal}
\begin{equation}
A_{\Lambda}\equiv \frac{\alpha_{\Lambda}+
\alpha_{\overline{\Lambda}}}{\alpha_{\Lambda}-
\alpha_{\overline{\Lambda}}}\,,
\end{equation}
where $\alpha_\Lambda$ ($\alpha_{\overline{\Lambda}}$) characterizes the
$\Lambda$ ($\overline{\Lambda}$) decay to  (anti)proton plus charged pion and,
if {\em CP} is a good symmetry in hyperon decay, $\alpha_\Lambda =
-\alpha_{\overline{\Lambda}}$. 

Fermilab E756~\cite{E756} and CLEO~\cite{CLEO} employed
a new technique in which the cascade decay of charged $\Xi$
hyperons is used to produce polarized $\Lambda$s, in whose subsequent decay the
slope of the (anti)proton angular distribution in the ``helicity" frame 
measures the product of $\alpha_\Xi$ and $\alpha_\Lambda$. If {\em
CP} is a good symmetry in hyperon decay this product should be identical for $\Xi$ and
$\overline{\Xi}$ events. The {\em CP}-asymmetry parameter measured is thus 
\begin{equation}
A_{\Xi\Lambda}\equiv \frac{\alpha_{\Xi}\alpha_{\Lambda}-
\alpha_{\overline{\Xi}}\alpha_{\overline{\Lambda}}}{\alpha_{\Xi}\alpha_{\Lambda}+
\alpha_{\overline{\Xi}}\alpha_{\overline{\Lambda}}}
\approx A_\Xi + A_\Lambda\,.
\end{equation}
The power of this technique derives from the large $\alpha$ value for the
$\Xi\to\Lambda\pi$ decay ($\alpha=0.64$). 
A further advantage in the fixed-target case is that within a given
${}^{{}^(}\overline{\Xi}{}^{{}^)}$ momentum bin the acceptances and
efficiencies for $\Xi$
and $\overline{\Xi}$ decays are very similar, since the switch from detecting
$\Xi$ to detecting $\overline{\Xi}$ is made by reversing the polarities of the
magnets, making the spatial distributions of decay products across the detector
apertures almost identical for $\Xi$ and for $\overline{\Xi}$. 
(There are still residual systematic uncertainties
arising from the differing cross sections for $p$ and $\overline{p}$ and for
$\pi^+$ and $\pi^-$ to interact in the material of the spectrometer.)

\begin{figure}[tb]
\vspace{-1.85in}
\centerline{\epsfysize 8in\epsffile{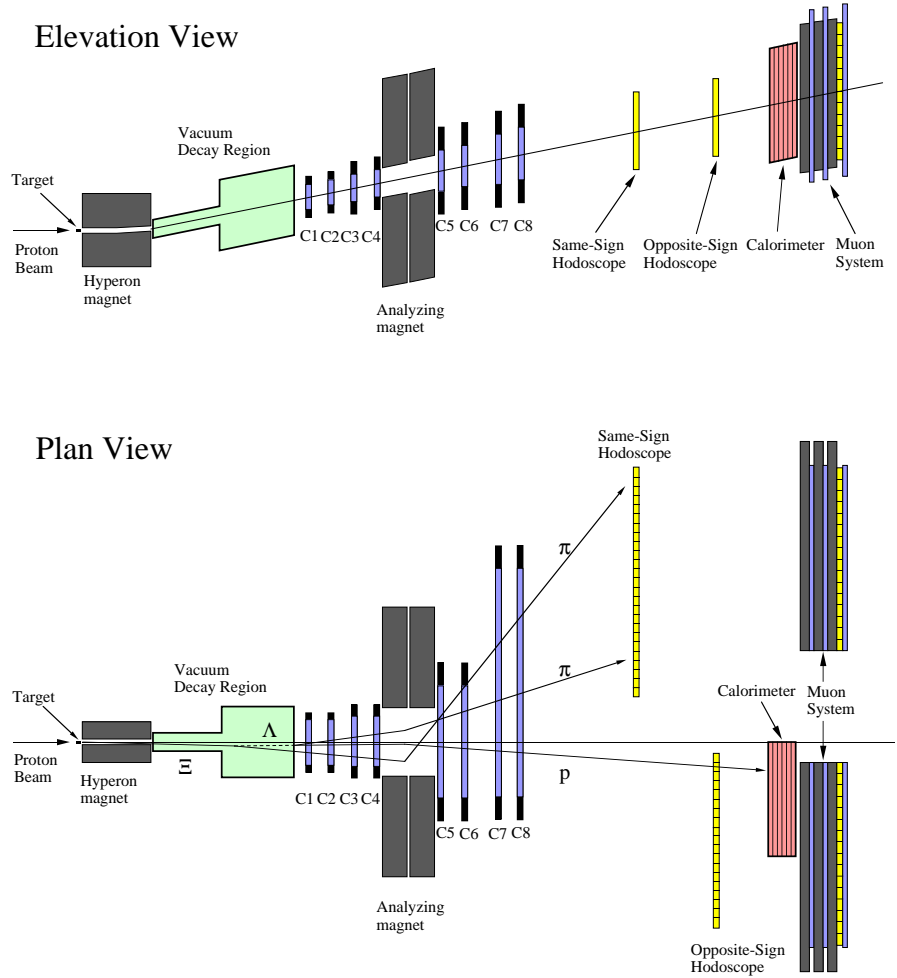}}
\vspace{-1.35in}
\caption {Schematic of the HyperCP spectrometer, comprising
eight (or in the 1999 run, nine) MWPC stations (C1--C8) located
downstream of the hyperon channel and decay pipe and surrounding an
analyzing magnet. The trigger calorimeter and hodoscopes are
located far downstream of the analyzing magnet, where the hyperon decay products have separated from each other and from the
beam, allowing the trigger elements to be kept outside the $\approx$20\,MHz
charged secondary beam. The muon detectors  give
sensitivity to rare decays.}
\vspace{-.25in}
\label{fig:HyperCP}
\end{figure}

Subsequent to E756, this technique has been used in the ``HyperCP"
experiment~\cite{E871} (Fermilab E871), depicted schematically in
Fig.~\ref{fig:HyperCP}, which ran during 1996--99.   Like E756, HyperCP used a
secondary charged beam produced by primary protons interacting in a metal
target. The secondary beam was  momentum- and sign-selected by means of a
curved 
collimator located within a 6-m-long dipole magnet. No measurements were
made until after the 13-m-long (evacuated) decay region. HyperCP recorded the
world's largest samples of decays of the $\Xi^-$ and $\overline{\Xi}{}^+$,
amounting to $2 \times 10^9$ and $0.5 \times 10^9$ events, respectively.   
When the analysis is complete, 
these should determine $A_{\Xi\Lambda}$ with a statistical uncertainty 
\begin{equation}
\delta A = \frac{1}{2\alpha_{\Xi}\alpha_{\Lambda}}
\sqrt{\frac{3}{N_{\Xi^-}}+\frac{3}{N_{\overline{\Xi}{}^+}}} = 1.4\times10^{-4}\,.
\end{equation} 
The Standard Model predicts this
asymmetry to be of order $10^{-5}$~\cite{ACP}.   
Thus if HyperCP sees a significant effect, it will be evidence for {\em CP}
violation in the baryon sector substantially larger than predicted by the
Standard Model.

\begin{table}
\caption {Summary of experimental limits on {\em CP} violation in hyperon decay.}
\label{tab:HCP}
\begin{center}
\begin{tabular}{lccc}
\hline
Experiment & Facility & Mode & $A_\Lambda$ [$^*$] or $A_{\Xi\Lambda}$ [$^\dagger$] \\
\hline\hline
R608 & ISR & $pp\to\Lambda X, pp\to\overline{\Lambda} X$ &
$-0.02\pm0.14^*$ \\
DM2 & Orsay & $e^+e^- \to J/\psi \to \Lambda\overline{\Lambda}$ & 
$0.01\pm0.10^*$
\\
PS185 & LEAR & $\overline{p}p\to\overline{\Lambda}\Lambda$ & 
$0.006\pm0.015^*$ \\
&  & $pN\to\Xi^- X, \Xi^-\to\Lambda\pi^-$, &  \\
\raisebox{1.5ex}[0pt]{E756} & 
\raisebox{1.5ex}[0pt]{Fermilab} &
$pN\to\overline{\Xi}{}^+ X, \overline{\Xi}{}^+\to\overline{\Lambda}\pi^+$ & 
\raisebox{1.5ex}[0pt]{$0.012
\pm0.014^\dagger$} \\
& & 
$e^+e^-\to\Xi^- X, \Xi^-\to\Lambda\pi^-,$ &\\
\raisebox{1.5ex}[0pt]{CLEO} & \raisebox{1.5ex}[0pt]{CESR} & 
$e^+e^-\to\overline{\Xi}{}^+ X, \overline{\Xi}{}^+\to\overline{\Lambda}\pi^+$ & 
\raisebox{1.5ex}[0pt]{$-0.057\pm0.064\pm
0.039^\dagger$}\\ 
\hline
\end{tabular}
\end{center}
\end{table}
\subsection{A future experiment}

Whether or not HyperCP observes a statistically-significant effect, it is of
interest to ask whether an experiment with substantially larger event samples
is feasible~\cite{Hyperon99}.  Since HyperCP sensitivity is an order of
magnitude short of the Standard Model prediction, a desirable goal would be two
orders of magnitude in sample size.

We have begun to explore this question.  
While we believe that the approach
taken in HyperCP is near the limit of what is possible with present-day
particle-detection technology,\footnote{The high rate of secondary beam in 
HyperCP ---
about 20\,MHz spread over an area of several cm$^2$ --- caused detector
inefficiencies in the beam region at the percent level (in the most upstream
MWPCs) due to MWPC deadtime.}
an alternative approach pioneered by the PS185 Collaboration at CERN may have
the requisite ``head room.\@"  The PS185 experiment~\cite{Barnes}  operated at
the Low-Energy Antiproton Ring (LEAR) at CERN between 1984 and 1996 and
utilized $\overline{p}p$ annihilation slightly above the threshold for
production of a  ${\overline\Lambda}\Lambda$ pair.  (In this case, the
requirement that the hyperon and antihyperon polarizations be precisely known
is modified, since by $C$-parity conservation in the strong interaction the
polarizations of the hyperons and antihyperons are equal.\footnote{This
assumption is in fact not experimentally tested at the level required for
this measurement, but the limit 
$B(\pi^0\to3\gamma)<3.1\times10^{-8}$~\cite{pi03g} suggests that it is a good
one~\cite{Pakvasa}; in any case, $C$ violation in the strong production process
would be as interesting a discovery as {\em CP} violation in the weak
decay~\cite{CP-Hyperon}.})

\begin{figure}[tbp]
\vspace{-1.85in}
\centerline{\epsfxsize 5 in\epsffile{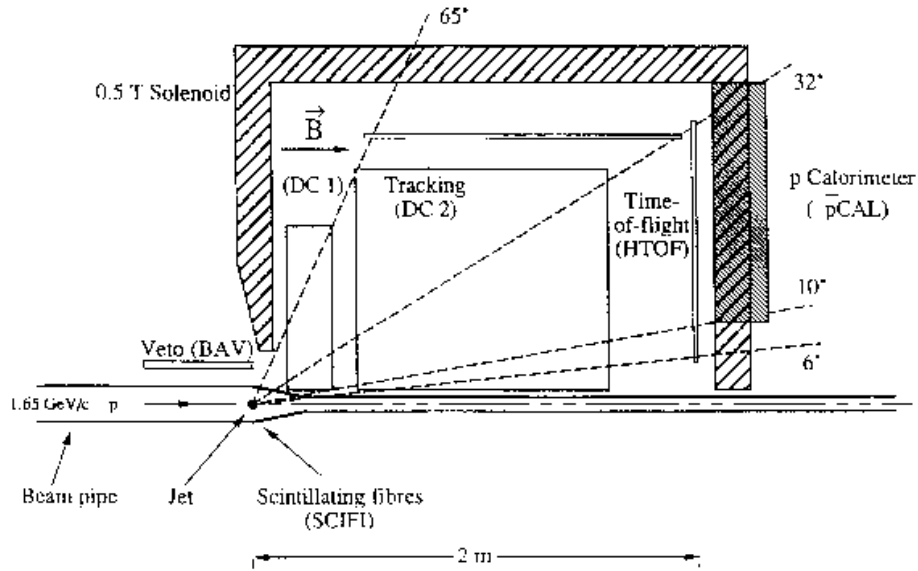}}
\vspace{-2in}
\caption {Schematic of ${\overline\Lambda}\Lambda$ spectrometer designed for 
SuperLEAR (from Ref.~\cite{CP-Hyperon}).}
\label{fig:SuperLEAR}
\end{figure}

Limited by the available antiproton intensity at LEAR, PS185 has achieved a
sensitivity of only  1.5\%~\cite{PS185}.  However, in the early 1990s the CERN
``CP-Hyperon Study Group" designed a hyperon {\em CP}-violation experiment for
SuperLEAR aimed at $10^{-4}$ sensitivity~\cite{CP-Hyperon} (see
Fig.~\ref{fig:SuperLEAR}). While SuperLEAR was never built, the antiproton
production rate at the Antiproton Source at Fermilab is already at least four
orders of magnitude beyond that achieved at LEAR, and, as mentioned above, 
substantial improvements
to its capabilities are planned.   A new antiproton storage ring at Fermilab
capable of producing  ${\overline\Lambda}\Lambda$ events at a 60\,kHz rate may
be feasible at relatively modest cost~\cite{Jackson}.  This would allow the
accumulation of a sample of order $10^{11}$ good events within a few years'
running time~\cite{Hyperon99}.  Challenges that will need to be met include the design of beam
optics and a gas-jet target that permit $\approx 10^{33}\,$cm$^{-2}$s$^{-1}$
luminosity, detecting the $\Lambda$ decay products and reconstructing their
tracks with  good efficiency at $\approx$200\,MHz charged-particle rate, triggering
with good efficiency and adequate background rejection at  $\approx$100\,MHz
interaction rate, and acquiring data at the resulting high trigger 
rate~\cite{pbar2000-Kaplan}.

\subsection{Additional physics}

Beyond hyperon {\em CP} violation, a low-energy ($p\approx2\,$GeV/$c$),
high-intensity antiproton storage ring equipped with a high-luminosity
fixed-target spectrometer may be able to carry out experiments designed to
study rare hyperon decays, hyperon beta decays, quark confinement, and soft QCD
effects~\cite{pbar2000}. A ``universal" fixed-target $\overline{p}p$
spectrometer is under study~\cite{GSI-upgrade}. Studies of charmonium and
possible gluonic-hybrid charmonium states would require higher antiproton
energies ($p\,{}_\sim$\llap{${}^<$}\,10\,GeV/$c$). Experiments in this energy
regime could address several open issues related to charmonium and QCD,
including the relative $J/\psi$ and $\psi^\prime$ decay  widths, $\chi_c$
widths, masses of the charmonium pseudoscalar states, the value of the strong
coupling constant $\alpha_s$ at the charm-quark mass, existence and properties
of higher radial excitations of charmonium, and the $J/\psi$-nucleon cross
section in the kinematic regime relevant to the interpretation of heavy-ion
quark-gluon-plasma searches~\cite{Seth}. Additional topics include the possible
existence of glueballs or gluonic hybrids in the $>2.5$\,GeV/$c^2$ mass region
(where their interpretation may be clearer than at low mass) and precision
measurements of the $\tau$ and $\nu_\tau$ masses~\cite{Seth}.
Experiments with trapped antiprotons include a variety of
{\em CPT} tests, for example precision studies of the hyperfine structure of
antihydrogen and of the gravitational interaction between antihydrogen and the
Earth~\cite{Holzscheiter,Phillips}. These might best be pursued in an
atomic-beam approach rather than at rest, thus a storage ring may have
advantages over the pulsed beam provided by the AD~\cite{Holzscheiter}.

\section{A SCENARIO FOR A NEW FERMILAB FACILITY}

Jackson has outlined a scenario leading to a new low-energy antiproton
storage-ring facility at Fermilab~\cite{Jackson,hbar}. The proposed approach is
to use the Main Injector to decelerate cooled antiprotons from the accumulator
down to the injection momentum of a new small storage ring ($\approx$2\,GeV/$c$).
While this is below the Main Injector's design momentum range, the ability of
the Main Injector magnets and their power supplies to operate stably at this
momentum has already been demonstrated, and a suitable operating mode for the
RF system has been devised and demonstrated to decelerate a proton
beam down to 3\,GeV/$c$~\cite{Jackson-private}. (In the Oct.\ 2000 test, a
software limitation prevented the demonstration of deceleration to 2\,GeV/$c$,
but this is expected to be rectified soon so that a test of deceleration to
2\,GeV/$c$ will become feasible~\cite{Jackson-private}.)

If the Main Injector indeed can provide 2\,GeV/$c$ antiprotons, the next steps
require construction of an extraction system for the decelerated antiprotons
followed by a transfer line to a location (say the MI-8 service area)
convenient for installation and operation of experimental trapping apparatus.
At this stage a low intensity of trapped antiprotons could be provided, using a
degrader to lower the momentum sufficiently for trapping.

To provide a higher-quality and higher-intensity beam, the 2\,GeV/$c$ ring will
be needed. It should be equipped with enough RF to decelerate down to the tens
to hundreds of MeV/$c$ range. With electron cooling in this ring, very high
luminosities ($\sim 10^{33}$) should be feasible. At this stage, hyperon
experiments, etc.\ (as discussed above) become feasible, provided that
sufficient antiproton flux is available to satisfy their needs. Interleaved
operation of other lower-flux experiments, including trapped-$\overline{p}$
studies using degraders or RFQ deceleration, should be possible with only
minor impact on the duty factor of the ``major user."

To provide a maximum of flexibility, an additional, larger storage ring could
be added to the facility to provide antiprotons over the
$\approx$1--10\,GeV/$c$ range, allowing both experiments at higher momenta and
periodic filling of traps with minimal impact on the high-luminosity
rare-hyperon-effect studies. However, whether such a ring would be worth
installing at Fermilab would depend on how the GSI upgrade
proposal~\cite{GSI-upgrade,Henning} fares.

The scenario just presented is being pursued aggressively by a private company,
Technanogy, LLC, with funding from NASA, and the small storage ring could be in
operation within the next several years.  It will soon behoove potential users
of the facility to work seriously on the design of the challenging experimental
apparatus that will be required, as well as to specify needed machine
capabilities so that these may be folded into the design process.

\section{Acknowledgements}

I thank the organizers for inviting me to speak at this stimulating conference,
convened at so striking a location.
This work was supported in part by grants from the Illinois Board of Higher
Education, the Illinois Department of Commerce and Community Affairs, and the
U.S. Department of Energy.

\end{document}